\documentclass[preprint,showpacs,preprintnumbers,amsmath,amssymb]{revtex4}
\usepackage{epsfig}
\preprint{JOPT/2004}
\begin{document}
\title{Reflectivity measurements in uniaxial superconductors: a methodological discussion applied to the case of La$_{2-x}$Sr$_{x}$CuO$_4$}
\author{M. Ortolani, P. Calvani, S. Lupi, and P. Maselli}
\affiliation{"Coherentia" - INFM and Dipartimento di Fisica, Universit\`a di Roma La Sapienza, Piazzale Aldo Moro 2, I-00185 Roma, Italy}
\author{M. Fujita and K. Yamada}
\affiliation {Institute for Chemical Research, Kyoto University, Gokasho,
Uji 610-0011, Japan}
\date{\today}

\begin{abstract}
Most of the novel superconductors are uniaxial crystals, with metallic planes ($ab$) orthogonal to an insulating axis ($c$). Far-infrared measurements of the reflectivity $R_{ab} (\omega)$ provide valuable information on their low-energy electrodynamics, but involve delicate experimental issues. Two of them are a possible contamination of $R_{ab} (\omega)$ from the $c$ axis and the extrapolation of the $R_{ab}$ data to $\omega$ =0, both above and below $T_c$. Here we discuss quantitatively these issues with particular regard to La$_{2-x}$Sr$_x$CuO$_{4+y}$, one of the most studied high-$T_c$ materials.

\end{abstract}
\pacs{74.25.Gz, 74.72.-h, 74.25.Kc}
\maketitle

\section{Introduction}

Infrared reflectivity measurements on single crystals have been largely employed to investigate the low-energy electrodynamics of novel superconductors, like high-$T_c$ materials, MgB$_2$, or Na$_x$CoO$_2$. However, the contradictory results often appeared in the literature show that such measurements involve delicate issues that will be discussed here. Among them, the most important ones are the possible contamination of the optical response of the conducting planes from the insulating axis, and the extrapolation of the reflectivity to $\omega=0$ both in the normal and in the superconducting phase.  
An example of the consequences of those issues is the present discussion, in the literature of high-$T_c$ superconductors, of their optical response in the far infrared. Indeed, depending on the existence or not of features peaked at frequencies low but not null, quite different models can be applied to the still unexplained transport properties of the $ab$ (Cu-O) planes. In the normal phase, experiments where the far-infrared conductivity $\sigma_1^{ab} (\omega)$ is smoothly decreasing as $\omega^{-1}$  lead to one-component approaches, like the "anomalous", or generalized, Drude model where both the scattering rate $\tau^{-1}$ and the effective mass are functions of $\omega$.\cite{Timusk}  On the other hand, the observation of resonances, \textit{i. e.}, peaks at finite frequencies in the far-infrared (FIR) lead to charges partially localized because of disorder, or phase separation, or ordering.\cite{Lobo, Startseva,Lupi2000,Venturini,Lucarelli,Takenaka,Dumm02, Bernhard} Such resonances, that have been reported most often for La$_{2-x}$Sr$_x$CuO$_{4-y}$ (LSCO), may lead the above models to unphysical predictions, like a change of sign in the carrier effective mass across the peak frequency.\cite{Lupi2004} 

In an attempt to explain such a complicated experimental landscape, it has been argued that most of the extra-Drude peaks observed in the $\sigma_1^{ab} (\omega)$ of LSCO are spurious\cite{Pimenov,comment} and caused by some leakage of the insulating $c$-axis response into that of the conducting $ab$ plane. Such leakage may be due either to polarizer inefficiency, or to a miscut, or bad polishing, or mosaicity of the crystals, usually grown by the Travelling-Solvent Floating-Zone (TSFZ) method. 

Flux-grown crystals, whose surface is intrinsically $ab$, should not have the above problems. However, FIR resonances quite similar to those reported for the TSFZ La$_{2-x}$Sr$_x$CuO$_{4+y}$  have been observed in\cite{Lobo} flux-grown La$_{2}$CuO$_{4+y}$ and in\cite{Bernhard} YBa$_{2}$Cu$_3$O$^{16,18}_{7-y}$ (YBCO). Unfortunately, flux-grown crystals are sometimes too small to provide reliable data at the long wavelengths where the resonances are reported to appear. 
Thin films are large, but their optical response is affected by the optical contribution of the substrate. This is subtracted through fitting procedures, which depend upon a number of parameters and may affect the results.  
In conclusion, identifying a correct procedure for measuring the reflectivity of TSFZ samples is of crucial importance for understanding the optical properties of LSCO and of other anisotropic materials. In the next Section we discuss this issue more extensively than usually found in the literature, and on a quantitative basis. We thus identify a procedure compatible with the required sensitivity, which will be applied to measurements, above and below $T_c$, on a TSFZ crystal of LSCO with $x$ = 0.12. 

\section{Measuring the reflectivity of an uniaxial crystal}

\begin{figure} 
{\hbox{\psfig{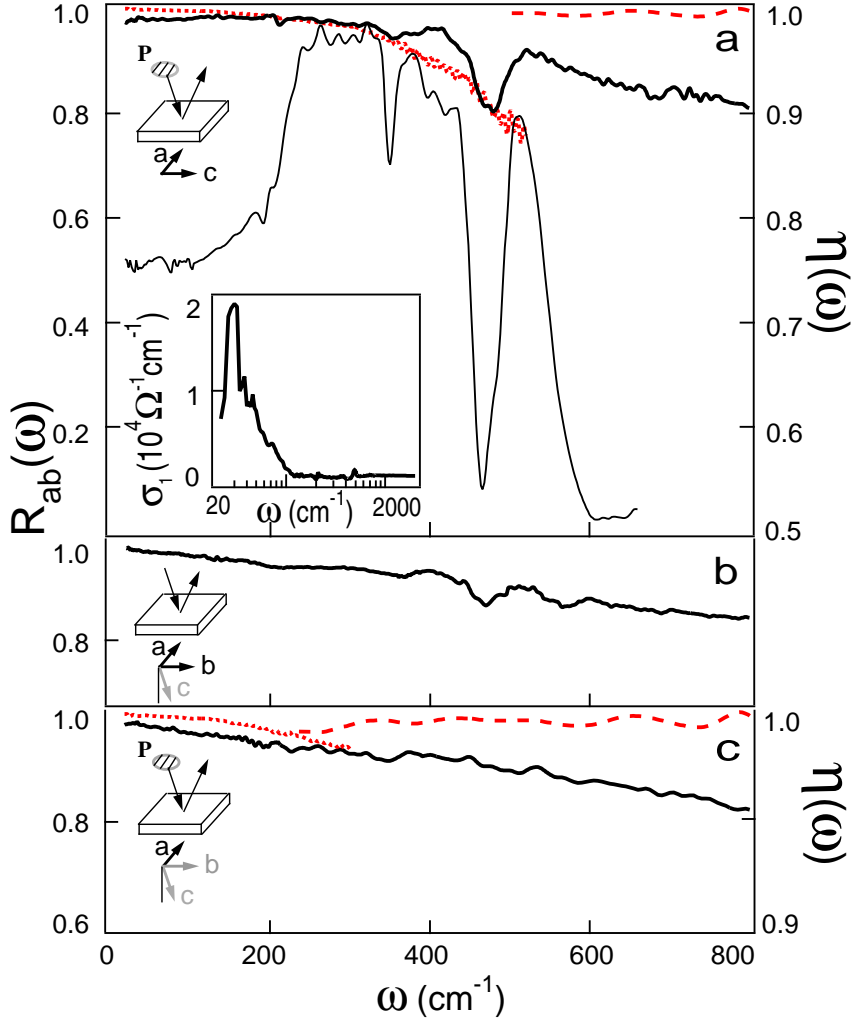}}}
\caption{Reflectivity, in the normal phase at 50 K, of La$_{1.88}$Sr$_{0.12}$CuO$_4$ obtained with different experimental configurations. In a), $R_{ab}$ (thick solid line, from Ref. \onlinecite{Lucarelli}) and $R_{c}$ (thin solid line, present work) as measured on the $ac (bc)$ surface by use of a polyethylene and a KRS-5 polarizer. Their efficiency $\eta$ is reported on the right scale by a dotted line and a dashed line, respectively, in the ranges where they were employed. The dip in $R_{ab}$ from the $c$ axis is due to a reduced efficiency of the polyethylene device above 300 cm$^{-1}$, which does not affect the observation of the peak in $\sigma_1^{ab}(\omega)$ reported in Ref. \onlinecite{Lucarelli}, and shown in the inset. In b), $R_{ab}$ is measured on the $ab$ surface of the same crystal as in a), by unpolarized radiation. A $\theta \simeq 1^{o}$ miscut, as determined by the procedure of Fig. 2, causes a slight $c$-axis leakage. In c), $R_{ab}$ is measured on the same $ab$ face as in b), but with the radiation field orthogonal to the miscut and by using the polyethylene polarizer below 250 cm$^{-1}$.}
\label{reflec}
\end{figure}

From the optical point of view, LSCO is a uniaxial crystal. Its dielectric function measured with the radiation field $\vec {E}$ along the $c$ axis, $\tilde{\epsilon}_{c}$, is insulating-like while that measured in the plane, $\tilde{\epsilon}_{ab}$, is metallic-like for $x > 0.055$. Those quantities are usually extracted from the reflectivity at quasi-normal incidence $R_{c}(\omega)$, and $R_{ab}(\omega)$, respectively, by Kramers-Kronig transformations. The strong anisotropy of the crystal may cause systematic errors that we analyze quantitatively in the present Section. We will refer to data taken in the normal phase at low temperature, where such errors are enhanced due to both the high $ab$ plane reflectivity, and the narrowing of the $c$-axis (phonon) absorption.
The top panel of Fig.\ \ref{reflec} shows $R_{ab}(\omega)$
for a single crystal of La$_{1.88}$Sr$_{0.12}$CuO$_4$ , grown by the TSFZ technique and accurately controlled, as it was reported in Ref. \onlinecite{Lucarelli}. $R_{ab}(\omega)$ was measured on the $ac(bc)$ surface of the crystal by use of commercial polarizers (made of polyethylene for the FIR and of KRS-5 for the mid infrared) in order to align $\vec {E}$ along $a(b)$. $R_{c}(\omega)$, measured afterwards on the same sample, is also shown. 

The resulting $\sigma_1^{ab}(\omega)$, reported in Ref. \onlinecite{Lucarelli} and recalled in the inset, showed a strong peak around 30 cm$^{-1}$ due to a change in the slope of $R_{ab}(\omega)$ slightly above that frequency. However, a dip appears in $R_{ab}(\omega)$ at 470 cm$^{-1}$, which indicates a "leakage" of the $c$-axis in the $ab$ response at that frequency. It has been argued\cite{comment} that the peak at 30 cm$^{-1}$ is also spurious as probably due to the same leakage. A check on the efficiency of the commercial polarizers employed in the experiment of Ref. \onlinecite{Lucarelli} shows that this is not the case. Indeed, the lines in Fig.\ \ref{reflec} (a) report the efficiency of the polyethylene and KRS-5 polarizer employed, $\eta = (I_{p}-I_{u})/(I_{p}+I_{u})$.  Here, $I_{p}$ ($I_{u}$) is the intensity of the field component parallel (orthogonal) to the polarizing direction. In correspondence with the dip in $R_{ab}$, the efficiency of the polyethylene device is lower than at 30 cm$^{-1}$. The error on $R_{ab}(\omega)$ due to $\eta < 1$ can be expressed in terms of the misalignment $\chi$ between the electric field of the radiation and the $a$ ($b$) axis by 

\begin{equation}
\chi \simeq  \tan \chi = \sqrt {I_{u}/I_{p}} \simeq \sqrt{(1-\eta)/2}
\label{chi}
\end{equation}

\noindent
Correspondingly, the real part of the measured dielectric function is
\begin{equation}
\epsilon_1^{eff} (\chi) = \epsilon_1^{ab} \cos \chi + \epsilon_1^{c} \sin \chi \simeq \epsilon_1^{ab} + \epsilon_1^{c} \chi = \epsilon_1^{ab} + \Delta \epsilon_1^{ab}
\label{epsilon-chi}
\end{equation}

\noindent
By applying the linear Kramers-Kronig transformations to both members of the last equality, recalling that $\epsilon_2 = (4\pi /\omega) \sigma_1(\omega)$, and evaluating the relative error on $\sigma_1^{ab}$ as in Ref. \onlinecite{Tanner93},
one obtains

\begin{equation}
\frac{\epsilon_2^{c} \chi}{\epsilon_2^{ab}} = \frac{\Delta \epsilon_2^{ab}(\omega, \chi)}{\epsilon_2^{ab}} = \frac{\Delta \sigma_1^{ab}}{\sigma_1^{ab}}  \simeq \frac{\Delta R_{ab}}{R_{ab}}\frac{R_{ab}}{1-R^2_{ab}} 
\label{error_sigma_chi}
\end{equation}

\noindent
Therefore, from Eq.\ \ref{chi} 

\begin{equation}
\frac {\Delta R_{ab}(\omega)}{R_{ab}(\omega)}\simeq  
\frac{1-R^2_{ab}(\omega)}{R_{ab}(\omega)} \sqrt{(1-\eta)/2} \frac{\epsilon_2^{c}}{\epsilon_2^{ab}} \, .
\label{error-R_chi}
\end{equation}

\noindent
We can now evaluate the error on the data of Fig.\ \ref{reflec}-a. At about 30 cm$^{-1}$, where the anomalous peak is observed in $\sigma_1 (\omega)$ and $\eta = 0.995 \pm 0.005$, the resulting $\Delta R_{ab}/R_{ab}$ is 0.1\%. As this systematic error is much lower than the noise on $R$, the measurement at low frequency is reliable, as it was pointed out in Ref. \onlinecite{reply}. On the other hand, $\eta$ rapidly drops to about 0.93 at 470 cm$^{-1}$, causing a 4\% deviation with respect to the real $R_{ab}$. Therefore, the latter spurious effect is not related with the observation of the resonance at 30 cm$^{-1}$. At higher frequencies, a KRS-5 polarizer with high and "flat" $\eta$ (see Figure) provides reliable data in the mid infrared.

However, usually $R_{ab}$ is measured by cutting and polishing the TSFZ sample along an $ab$ surface and using unpolarized radiation. In order to evaluate quantitatively the systematic error in this case, we cut the same La$_{1.88}$Sr$_{0.12}$CuO$_4$ parallel to the $ab$ planes within the mechanical precision of the diamond-wire saw, and we accurately polished the face. Then, we measured the X-ray rocking curve on a matrix of 3x3 points of the $ab$ surface. The width of the (0,0,2) diffraction peak was determined, and found to correspond to a mosaicity $\leq$ 0.2$^0$. Therefore, any possible source of error from the crystal quality was excluded. 

Anther systematic error may be related, however, to the average angle of incidence under which the radiation impinges on the crystal. It will depend on the geometry of the reflectivity set-up and on the angular width of the beam, namely on the $f$-number of the optical device. In the present case, the central ray had an incidence of 8$^0$ and the $f$-number was 4. Therefore the maximum angle under which the marginal ray hit the surface was about 15$^0$. Now, any deviation from normal incidence smaller than about $15^{o}$ does not affect the result in the FIR.\cite{Marel} Indeed, as the radiation enters the metallic cuprate, due to refraction it propagates very close to the $c$ axis. This misalignment may instead slightly affect $R_{ab}$ in the mid infrared, where the refraction index $n$ decreases.

\begin{figure}
{\hbox{\psfig{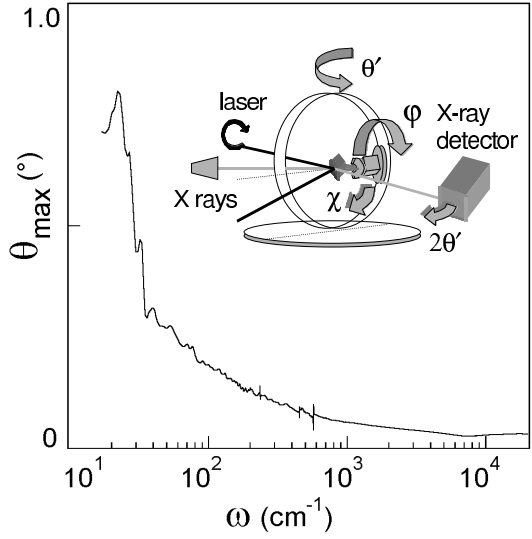}}}
\caption{Maximum misalignment $\theta_{max}$ between the electric field of the radiation and the $ab$ plane of a cuprate, that can be tolerated according to Eq.\ \ref{error-R}, as a function of $\omega$. The data of Fig. 1 were used in this example. The inset shows the setup used for measuring the miscut angle $\theta$. The sample is mounted on a four-circle diffractometer to align the $c$ axis in the $\chi$-circle. The normal to the surface, identified by the laser reflection, makes a precession under a sample rotation by $\phi$. By adjusting the sample holder until no precession is observed, the surface normal is brought in the $\chi$-circle. Then, by varying $\theta'$ until the Bragg peak is observed again, the miscut angle $\theta$ is determined.}
\label{theta}
\end{figure}

However, in Fig.\ \ref{reflec}-b $R_{ab}(\omega)$ still shows an appreciable imprint of the $c$ axis TO phonon, even if smaller than in (a). Indeed, by using the procedure described in the inset of Fig.\ \ref{theta}, we found out a miscut angle $\theta$ = 1$^o \pm 0.5^{o}$ between the crystal surface and the $ab$ planes of LSCO. The uncertainty is due to both surface roughness and laser beam divergence. 

We can now calculate the error $\Delta R_{ab}/R_{ab}$ that the small miscut $\theta$ induces on the measured reflectivity, and show how it can be reconducted within the experimental noise.
The response to unpolarized radiation which propagates with $\theta \neq 0$ with respect to the principal ($c$) axis of a uniaxial crystal is an effective dielectric function\cite{Born}

\begin{equation}
\epsilon_1^{eff} (\theta) = \frac{1}{2} \left[\epsilon_1^{ab} + \frac{1}{\frac
{\cos^{2}\theta}{\epsilon_1^{ab}
}+\frac{\sin^{2}\theta}{\epsilon_1^{c}}} \right] 
\simeq \epsilon_1^{ab} - \frac{1}{2} (\epsilon_1^{ab} - \epsilon_1^{c})\frac{\epsilon_1^{ab}}{\epsilon_1^{c}} \theta^2 = \epsilon_1^{ab}(\omega, \theta) + \Delta \epsilon_1^{ab}(\omega, \theta) \, ,
\label{error_eps1}
\end{equation}

\noindent 
where the term in the square bracket has been expanded to the second order for small $\theta$. By using again the Kramers-Kronig transformations, one obtains

\begin{equation}
\frac{\Delta \epsilon_2^{ab}(\omega, \theta)}{\epsilon_2^{ab}} = \frac{\Delta \sigma_1^{ab}}{\sigma_1^{ab}}  = 
\theta^2 \left[\frac{-2 \omega}{\pi \epsilon_2^{ab}} \large \wp \int_0^\infty {\frac {-(\epsilon_1^{ab})^2/2\epsilon_1^{c}}{\omega'^2 - \omega^2} d\omega'}  + \frac{ 1}{2} \right] = \theta^2 g(\omega) \, .
\label{error_sigma}
\end{equation}

\noindent 
One wants that the systematic error due to the $c$-axis leakage is smaller than the standard uncertainty (noise) on $R_{ab}$ (here assumed to be $\pm$ 0.5\%). Therefore, by using again the criterion in Ref. \onlinecite{Tanner93},
one may write

\begin{equation}
\frac {\Delta R_{ab}(\omega)}{R_{ab}(\omega)}\simeq  \frac{1-R^2_{ab}(\omega)}{ R_{ab}(\omega)} \theta^2 g(\omega) < 0.005
\label{error-R}
\end{equation}

\noindent 
From the above formula one can extract the maximum misalignment to be tolerated, $\theta_{max} (\omega)$. By using at first order one of the data sets in Fig.\ \ref{reflec}, one obtains from Eq.\ \ref{error_sigma} the curve in Fig.\ \ref{theta} for our La$_{1.88}$Sr$_{0.12}$CuO$_4$ crystal. As one can see, it leads to a strict condition (miscut $\theta_{max} < 1^0$) that can be fulfilled only by the procedure described in the inset of Fig.\ \ref{reflec}-c. Therein, the surface is again $ab$, but the radiation is linearly polarized in a direction orthogonal to the plane of the miscut angle. Therefore, in the new configuration one has

\begin{equation}
\epsilon_1^{eff} (\chi, \theta) = \epsilon_1^{ab} \cos \chi + \sin \chi \frac{1}{\frac
{\cos^{2}\theta}{\epsilon_1^{ab}}+\frac{\sin^{2}\theta}{\epsilon_1{c}}} 
\label{epsilon-eff}
\end{equation}

\noindent
so that

\begin{equation}
\frac {\Delta R_{ab}(\omega)}{R_{ab}(\omega)}\simeq  2 \sin \chi \frac{1-R^2_{ab}(\omega)}{ R_{ab}(\omega)} \theta^2 g(\omega) < 0.005
\label{error-R_bis}
\end{equation}

\noindent
where $\chi$ is given by Eq.\ \ref{chi}, while $\theta$ and $g$ are the same as in Eq.\ \ref{error_sigma}. By using the polyethylene polarizer up to 250 cm$^{-1}$, and a KRS-5 polarizer at higher frequency, the effective deviation from an ideal $ab$ plane is reduced by the factor $2 sin \chi$ to $\sim 10^{-2} \theta^2$, fully compatible with the small error that can be tolerated on $R$. Indeed, as one can see in the bottom panel of Fig.\ \ref{reflec}, the dip at  470 cm$^{-1}$, which indicates a leakage with the $c$ axis, has disappeared. If, moreover, the plane of incidence is such as to coincide with the miscut plane ($s$ polarization) the misalignment due to a nonzero angle of incidence is also ruled out, for any refraction index $n$. This configuration was indeed selected here to obtain either the results of Fig.\ \ref{reflec}-c for $x$ = 0.12 and those of Fig.\ \ref{sigma}-c for $x$ = 0.26. Therein, full $R(\omega)$ data sets for both samples are shown in the inset. Between 15 cm$^{-1}$ and $\sim$ 12000 cm$^{-1}$ we took as reference a gold film and at  higher frequencies a silver film, both of which were evaporated \textit{in situ} onto the sample. Suitable corrections for the $R (\omega) <1$ of both metals were introduced. The final alignment of the focusing mirrors in the reflectivity set-up
was performed by use of remotely controlled motors after evacuating the interferometer. In this way, any misalignment due to mechanical
stress of the sample compartment could also be eliminated.

\section{From the reflectivity to the optical conductivity}

\begin{figure}
{\hbox{\psfig{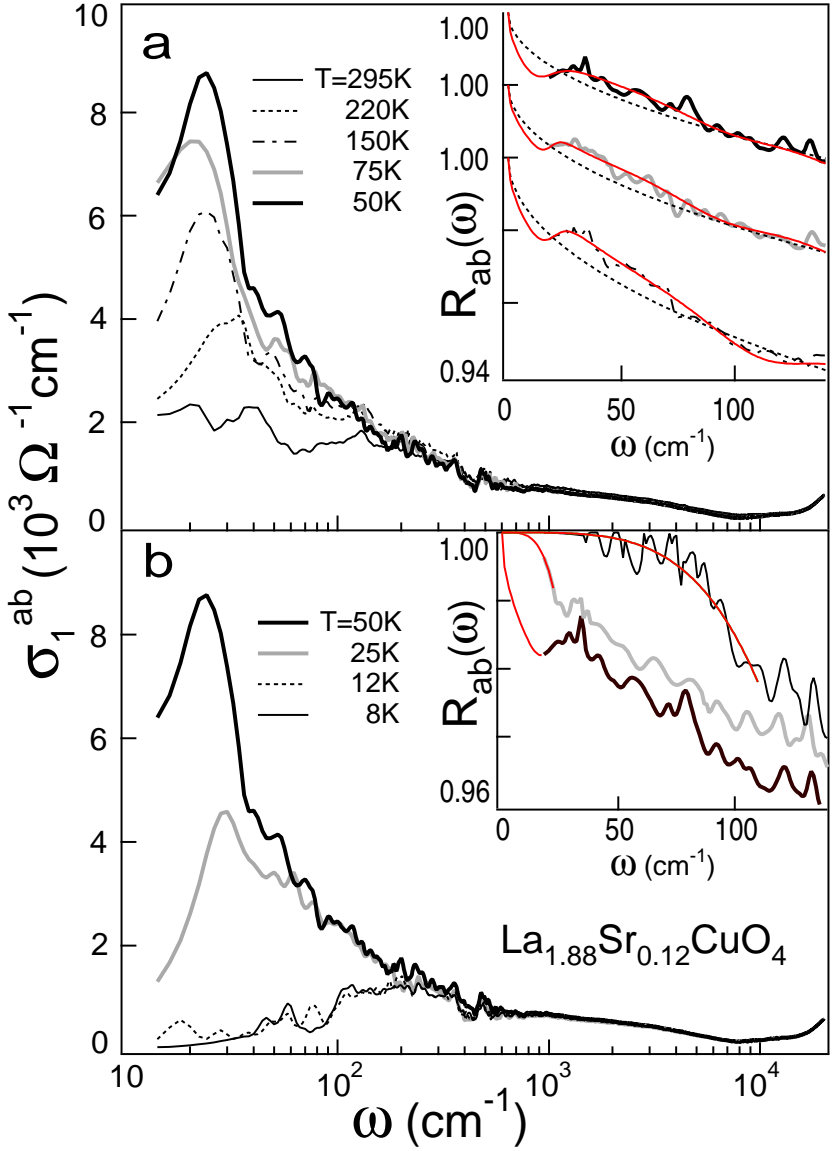}}}
\caption{Optical conductivity, from the reflectivity data reported in the inset, of the underdoped crystal $x$ = 0.12 above $T_c$ (a) and below $T_c$ (b). The insets show the corresponding $R_{ab}$: in the FIR above (a) and below $T_c$ (b). In the inset (a) the red lines are Drude-Lorentz fits used in the extrapolation in the normal phase. The dashed lines are Hagen-Rubens extrapolations for comparison. In the inset (b) the red lines are fits based on the London model (see text). } 
\label{sigma}
\end{figure}

The optical conductivity $\sigma_1^{ab} (\omega)$ was extracted from the reflectivity of Fig.\ \ref{reflec}-c by Kramers-Kronig transformations, by extrapolating $R_{ab} (\omega)$ to high frequencies as described in Ref. \onlinecite{Lucarelli}. The extrapolation to zero frequency deserves a more detailed discussion. In the normal phase above $T_c$, most authors use the Hagen-Rubens (HR) extrapolation 
$R_{ab} (\omega) = 1 - A \sqrt \omega$. This holds for $\omega < \Gamma$ in a conventional Drude absorption of width $\Gamma$, but may be not suitable to describe the reflectivity of the cuprates. Indeed, as shown in the inset of Fig.\ \ref{sigma}-a, at any $T \leq$ 220 K, the HR extrapolation (dashed line) does not fit the extra reflectivity observed in the FIR. Any attempt to include this latter in a different HR fit leads to a dc conductivity, higher by at least one order of magnitude than those reported in the literature.\cite{Ando} Therefore, we used at any $T$ a Drude-Lorentz fit to $R_{ab} (\omega)$ for $\omega \geq$ 15 cm$^{-1}$, which fits (see the inset of Fig.\ \ref{sigma}-a) the slight but reproducible decrease observed below 40 cm$^{-1}$ and provides a suitable extrapolation for $\omega \to 0$. It is our opinion that, in some previous studies, the choice \textit{a priori} of an HR law for $R_{ab} (\omega)$ may have prevented the detection of peaks at finite FIR frequencies in the $\sigma_1^{ab} (\omega)of $LSCO. 
At frequencies higher than 20000 cm$^{-1}$ we used the reflectivity data of Ref. \onlinecite{Tajima2}, that were extrapolated to infinite frequency by the power law  $R(\omega) \propto \omega^{-4}$.
Below $T_c$, different extrapolations to $\omega = 0$ can be used. In high-$T_c$ superconductors, it is well known that only a fraction of the in-gap spectral weight 

\begin{equation}
W (\Omega,T) = \int_0^\Omega {\sigma_1(\omega,T)d \omega}\, , 
\label{weight}
\end{equation}

condenses below $T_c$. In LSCO, such fraction is on the order of 20 \%.\cite{Tanner00} To keep track of this evidence, a two-fluid model is often employed, where the superfluid component is modeled by a Drude term with vanishingly small scattering rate, and the normal fraction is represented by a Lorentzian lineshape with finite width. However, it is also assumed in general that high-$T_c$ cuprates are in the clean limit, due to their extremely short coherence length. Therefore, the extrapolation law $R = 1 - A \omega^4$ is probably more suitable, as it derives from the London model which describes the electrodynamics of a superconductor with infinite gap. Indeed, we have made the latter choice in the present experiment. As one can see in the inset of Fig.\ \ref{sigma}-b, this law provides a good fit to data down to the lowest measuring frequency, with one parameter only. Both the above approaches are clearly phenomenological. An attempt to extend the BCS theory to any situation intermediate between the clean limit of the London assumption and the dirty limit of the Mattis-Bardeen approach ha been performed by Zimmermann.\cite{Zimmermann}

The optical conductivity $\sigma_1^{ab} (\omega)$ of the underdoped crystal $x$ = 0.12 is shown in Fig.\ \ref{sigma} for temperatures above (top) and below $T_c$ (middle),  and compared with that measured in a crystal with $x$ = 0.26 (bottom). In the normal phase, in the underdoped sample an anomalous FIR peak is still observed, despite any possible contribution from the $c$ axis has been ruled out by the procedure described in the preceding Section. That feature is very similar, both in peak frequency and temperature dependence, to the one measured in Ref \onlinecite{Lucarelli} in the $ac$ plane of the same $x$ = 0.12 sample, and extracted from the data of Fig.\ \ref{reflec}-a. Such anomaly, which is not observed in non-superconducting LSCO metals with $x$ = 0.26, was attributed to excitations of clusters of localized charges, possibly stripes, on the basis of phenomenological\cite{Lucarelli} and theoretical\cite{Benfatto} arguments.

\section{Conclusion}

In the present paper we have addressed an issue that is widely debated in the literature of optical spectroscopy of novel superconductors: a reliable determination of their far-infrared conductivity, in the presence of strong optical anisotropies. In cuprates, the discussion has regarded mainly the existence, or not, of resonances at low energy that could be related to carrier localization and phase separation. Slight differences in the $ab$ plane reflectivity and in their extrapolation to zero frequency, not much larger than the experimental noise, can lead to either conclusion. In MgB$_2$, the possibility of optically observing both energy gaps which are known to exist in this material will depend on the solution of the same issues. In the recently discovered "wet" superconductor, Na$_x$CoO$_2$.nH$_2$O, a major problem for the spectroscopist will be the observation, in the far infrared, of a tiny energy gap in the presence of water absorption. 
We have then proposed here a quantitative discussion of the two main problems, above mentioned, which affect the experiments in the far infrared. 
We have analyzed quantitatively the error that an admixture of the $c$-axis contribution, due to different reasons, may cause on the in-plane reflectivity of an uniaxial crystal like LSCO. We have then identified a safe procedure which has been shown to reconduct the unavoidable miscut of the crystal surface within the experimental noise. This is obtained by using a polarizer aligned orthogonally to the miiscut, which reduces the $c$-axis contribution by nearly two orders of magnitude. This method, applied to an $ab$ surface of the same LSCO crystal that was measured on the $ac$ side in a previous experiment, has confirmed the observation, reported therein, of a resonance in the far infrared.\cite{Lucarelli}
We have then discussed the extrapolation of the reflectivity to $\omega$ = 0, that is necessary to extract $\sigma_1^{ab} (\omega)$ from the raw data by Kramers-Krong transformations. We have pointed out that the standard Hagen-Rubens law used for superconductors in their normal phase may erase important details that, even if relatively small, would make major features appear in $\sigma_1$ when $R_{ab} (\omega)$ is close to 1. The use of extrapolations based on Drude-Lorentz fits would eliminate this risk.
Finally, we have addressed the same issue for the reflectivity below $T_c$, by comparing the advantages and disadvantages of different choices. In MgB$_2$ the BCS model has proved to provide accurate extrapolations. Concerning high-$T_c$ superconductors, the models proposed and used up too now are not fully satisfactory for some reason. There is certainly a need for more detailed models which may take into account the $d$-wave symmetry of the pair wavefunction in these materials.

\begin{acknowledgments}
We are indebted to M. Colapietro for performing in his laboratory most of the X-ray diffraction measurements here used to characterize the sample.
\end{acknowledgments}

\end{document}